# MANAGING AND QUERYING WEB SERVICES COMMUNITIES: A SURVEY

Hela Limam[1] and Jalel Akaichi[2]

[1,2]Department of Computer Sciences, ISG,SOIE,University of Tunis, Tunis, Tunisia

[1]`hela.limam@isg.rnu.tn`
`2jalel.akaichi@isg.rnu.tn`

## ABSTRACT

*With the advance of Web Services technologies and the emergence of Web Services into the information space, tremendous opportunities for empowering users and organizations appear in various application domains including electronic commerce, travel, intelligence information gathering and analysis, health care, digital government, etc.*
*However, the technology to organize, search, integrate these Web Services has not kept pace with the rapid growth of the available information space. The number of Web Services to be integrated may be large and continuously changing.*
*To ease and improve the process of Web services discovery in an open environment like the Internet, it is suggested to gather similar Web services into groups known as communities.*
*Although Web services are intensively investigated, the community management issues have not been addressed yet*
*In this paper we draw an overview of several Web services Communities' management approaches based on some currently existing communities platforms and frameworks. We also discuss different approaches for querying and selecting Web services under the umbrella of Web services communities' .We compare the current approaches among each others with respect to some key requirements.*

## 1. INTRODUCTION

Nowadays with the incredible growth of the information space, the hard competition between enterprises populating it, competition is not limited to goods, services or software products but also includes Web services . Web services are gaining momentum as a way to interact applications across organizations. The increasing number of available Web services, the growing need to collaborate and to share knowledge guiding to better decisions are factors which generate a growing interest for gathering Web services into communities in order to speed up and facilitate Web services discovery and selection.

The emergence of Web Services Communities as a model for integrating heterogeneous web information has opened up new possibilities of interaction and offered more potential for interoperability. In fact, gathering Web services into communities aims to address complex users' needs that a single Web Service can not satisfy. Web services communities provide a centralized access to several functionally-equivalent Web services via a unique endpoint which enables the query processing.

However, the organization into communities raises management issues: how to initiate, set up, and specify a community of Web services. These issues have been addressed in different ways

DOI: 10.5121/ijdms.2011.3107        93

according to several approaches. In this work we mainly aim at presenting the concepts and operations that are required to specify and manage a community of Web services according to different approaches. We draw a survey on existing proposals for organizing Web services into communities.

These communities are built with the purpose to be queried transparently and easily by users, which aim to satisfy their informational needs in a satisfactory time and in a pertinent retrieval. In fact, a user query may involve the access of a number of distributed communities in order to locate Web services that are capable of answering the query which is not locally available. Different approaches tackle the query processing among communities following different point of views. We draw an overview on them following some key requirements involved in this task. Finally, we study the web services selection process enabled by the organization into communities.

This paper is organized as follows: Section 2 introduces the notion of community for gathering Web services; section 3 summarizes communities' management frameworks and compares them among each others with respect to some key requirements. Section 4 discusses Web services communities' approaches for querying Web services communities and presents related work on Web services selection for Web services communities, and section 5 concludes our work and presents some insights for future work.

## 2. DEFINITION OF WEB SERVICES COMMUNITIES

A community has been defined as a group of people living together or united by shared social interactions, social ties, and a common 'space' [1]; as a social network of relationships that provide sociability support, information, and a sense of belonging [2], and as a set of relationships where people interact socially for mutual benefit [3].
In [4], a virtual community differs from other communities only in that its common space is cyberspace. Virtual communities therefore describe the union between individuals or organizations who share common values and interests using electronic media to communicate within a shared semantic space on regular basis.

The term community is not particular to Web services. In grid computing for example, solutions for sharing resources in a grid rely on communities [5].
The concept of e-catalog community is used as a way of organizing and integrating a potentially large number of dynamic e-catalogs [6]. An e-catalog community is a container of catalogs that offer products of a common domain (e.g., community of laptops). It provides a description of desired products without referring to any actual provider (e.g., Dell.com).

When applied to Web services, communities help gathering Web services that provide a common functionality in order to simplify the access to Web services via a unique communication endpoint, which is the access point to the community
In [7] authors develop Web Service Community (hereafter WSC) as a promising computational infrastructure that facilitates universal description of service capability, allowing automated dynamic selection of the best service.

In this paper, Web Service Community (WSC) is proposed as a promising computational infrastructure in which web service is described in the way of concordant combination of explicit representation and implicit representation and attention on both commonness and peculiarity of service individual.

Benatallah et al. define [8], a community as a collection of Web services with a common functionality, although these Web services have distinct non-functional properties like different providers and different QoS parameters The concept of service community is a solution to the problem of composing a potentially large number of dynamic Web services. A community describes the capabilities of a desired service without referring to any actual Web service providers. In other words, a community defines a request for a service which makes abstraction of the underlying providers

In [9], the concept of community gathers services from the same domain of interest and publishes the functionalities offered by Web services as generic operations. The authors provide a general template referred to as community ontology for describing semantic Web services and communities. A community is a "container" that clusters Web services based on a specific area of interest (e.g., disability, adoption Communities provide descriptions of desired services (e.g., providing interfaces for insurance services) without referring to any actual service).All Web services that belong to a given community share the same area of interest.

## 3. MANAGING WEB SERVICES COMMUNITIES

Web services communities appear to be a solution towards reshaping online communication and collaboration between Web services. However, the organization into communities raises management issues: how to initiate, set up, and specify a community of Web services.

### 3.1. The management requirements

The main requirements related to communities' management can be resumed by creating and updating Web Services Communities then building relationships between them. These requirements are detailed in the following, and illustrated through an UML Use Case diagram in figure 1.

- Community Creation: The Community Manager creates a community by grouping Web services related to the same domain then he defines its schema to provide a description of the field to which the community belongs without referring the Web Services providers.
- Community Update: As communities evolve in the Web environment characterized by its dynamism, changes can frequently affect communities. Hence, communities should be permanently updated. The community update takes in general two forms, deletion or modification.
- Community deletion: The community that does not contain any Web Service is deleted, for this purpose the Community Manager has to identify a community that users constantly leave without performing any further action.
- Community modification: It consists on adding or updating members and consequently adding or updating Web Services manually or automatically.
- Building relationships between communities: Communities may create relationships between each others. Relationships between communities fall into two types: Peer relationship or specialization relationship.
    - Peer Relationship: To form peer relationship between communities, the Community Manager searches other communities whose domains are similar to their community.
    - Specialization Relationship: It represents specialization between two communities' domains (for example, Hospital is a sub-community of Medical Institutions).

- The Web Services providers' registration: In order to be accessible through a community, the Web Service provider must apply for registration. The registration of its Web Service is done by the Member Manager. When a request for registration arrives, the Member Manager associates the web service provider with the correspondent community. By registration, the Web Service provider becomes a member of the community.

- Converting Web Information Sources to Web Services: In case of Web information source does not take the form of Web Service, the Member Manager has to convert it to Web Service.

- Defining Communities Members: The Web Services provider defines communities members by feeding communities with Web Services related to its domain.

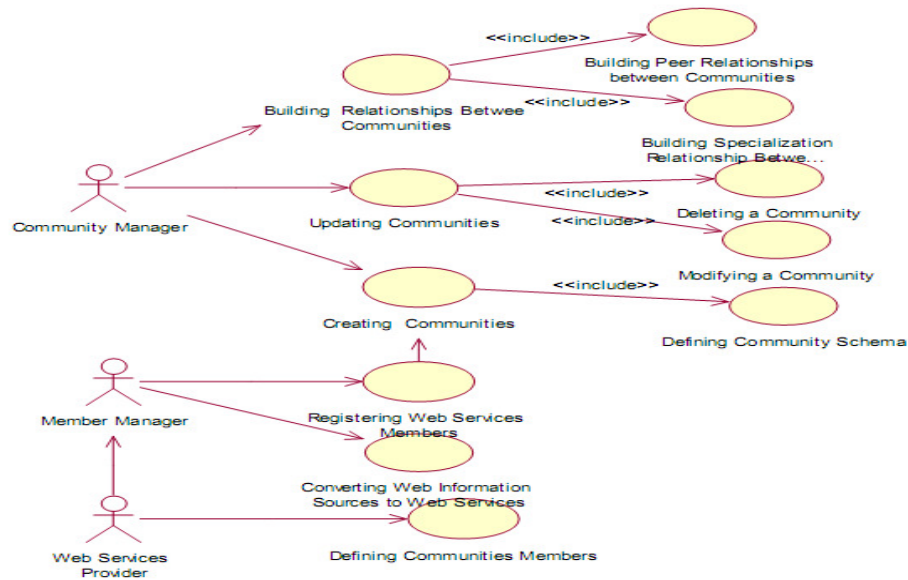

Figure 1. UML Use Case diagram for communities' management

### 3.2. Communities management: Current approaches

Several works gather functionally-similar Web services into communities that are accessed via a common interface and propose solutions for tackling Web services communities' management issue described above.

Such a solution is proposed in SELF-SERV framework [10], which distinguishes 3 types of services: elementary services, composite services, and service communities.

Elementary service might be a Web form-based interface to a weather information source. A composite service aggregates multiple Web services, which are referred to as its components. An example of a composite service would be a Web-accessible travel preparation service, integrating autonomous services for bookingflights, booking hotels, searching for attractions, etc. Service communities can be seen as containers of alternative services. Several mediators establish correspondences between the community interface and Web services that implement the functionality of the community.

In order to be accessible through communities, pre-existing Web services can register with them. Services can also leave and reinstate these communities at anytime. At runtime, when a community receives a request for executing an operation, it selects one of its current members, and delegates the request to it.

Whether elementary, composite, or community-based, a Web service is specified by an identifier (e.g., URL), a set of attributes, and a set of operations. The attributes of a service provide information which is useful for the service's potential consumers (e.g., public key certificates). In order to ensure that all services provide a uniform interface, each service in SELF-SERV is wrapped by a software component hosted by its provider. A service's wrapper acts as its entry point, in the sense that it handles requests for executing the operations provided by the service.

The SELF-SERV framework features are service manager and several pools of services. The service manager consists of a service discovery engine facilitating the advertisement and location of services, a service editor facilitating the definition of new services and the edition of existing ones, and a service deployer generating routing tables of every state of a service's state-chart and uploading these tables into the hosts of the corresponding composite service.

Service composition is based on state-charts, gluing together an operation's input- and output-parameters and produced events. Service execution is monitored by software components called coordinators, which initiate, control, and monitor the state of a composite service they are associated with. The coordinators retrieve the state relevant information from the service's state-chart and represent it in what is called a routing table containing pre-conditions and post-processings.

WS-catalogNet is a Web services based data sharing middleware infrastructure whose aims is to allow the integration of large number of e- catalogs. The concept of e-catalog community is used as a means to architect the organization and the integration of a potentially large number of dynamic e-catalogs [8].

WS-CatalogNet offers a set of tools which allow creating communities, registering e-catalog members, creating peer relationships between communities, querying individual communities and routing queries among communities.

An e-catalog community [8] is a container of catalogs that offer products of a common domain (e.g., community of laptops). It provides an ontological description of desired products without referring to any actual provider (e.g., Dell.com). Communities of e-catalogs are established through the sharing of high-level meta-information. Actual providers can register with any community of interest to offer the desired products. E-catalog providers can join or leave any community of interest at any time. . Catalog communities may newly form or disappear. Communities meaningfully organize and divide the information space into groups of manageable spaces (e.g., putting similar products together).

The fundamental element of a Web Service Community (WSC) according to [7] is subject-club which is something like a special service container in which some services localized in different place across the globe but with similar domain interest are clustered.

Subject-club ontology, one important part of the hierarchical community ontology, is divided into club local ontology and web service private ontology, respectively  characterizing commonness and peculiarity of services with similar function, because practice has proved that the description on commonness and peculiarity of web service deserve same considerations.

Obviously Web service description in WSC combines explicit representation and implicit representation concordantly, taking both commonness and peculiarity of web service into account. This prominent feature helps the service discovery and selection process to improve efficiency and flexibility by narrowing search space into certain one or several subject-clubs.
Subject-club and leading service list are two conspicuous components in the structure of WSC. Like business-to-service switch, subject-club acts as not only business function logic sorter but also web service container. Leading service list manifests directly the competition among member services of one subject-club and has outstanding capacity to respond to dynamic change of single service, especially to those services with better quality performance.

In [11], authors propose an approach that supports the concepts, architecture, operation and deployment of Web service communities. The notion of community serves as an intermediary layer to bind to Web services. A community gathers several slave Web services that provide the same functionality. The community is accessed via a unique master Web service. Users bind to the master Web service that transparently calls a slave in the community. . This work details the management tasks a master Web service is responsible for. Such tasks include among other things registering new Web services into the community, tracking bad Web services, and removing ineffective Web services from the community. A master Web service represents the community and handles users' requests with slave Web services with the help of a specific protocol.  The community is managed as the following:

- The master Web services send a call for bid to the slave Web services of the community.

- Slave Web services assess their current status and availability to fulfil the resquest of the master Web service, and interested Web service reply to the call.

- The master Web service examines the received proposals and chooses the best Web services according to its preferences (QoS, availability, cost, fairness. . . ). It notifies the winner slave Web service.

- Slave Web service that answered the call for bid but were not selected are notified too

Built upon this work, authors propose in [12] context-based semantic mediation architecture for Web service communities. Indeed, the applicability of the mediation proposition goes beyond this domain. However, they specifically focus  on its deployment with communities as defined in [11], where semantic mediation is performed between the community master and slave Web services.

The context-based model proposed has for objective to ease the task of Web service providers when they decide to adhere to new communities, by scaling domain ontologies down to the minimum, and providing additional context ontologies to handle the different local semantics of service providers.

The mediation architecture for Web service communities is built on a master Web service that contains a mediation module. This mediation module enables the master Web service to handle incoming requests from outside the community.

Thanks to the mediation module, the master Web service can act as a mediator. Upon reception of a user's request, it uses the mediation module to convert the message into the slave Web service's semantics. Upon reception of an answer from a slave Web service, the master Web service uses the mediation module again to convert the message into the semantics of the community before sending it back to the user. The master Web service is also responsible for other tasks, such as selecting a slave Web service upon reception of a request or managing the community.

A community-based architecture for semantic Web services is proposed in [13]. In this work, communities gather services from the same domain of interest and publish the functionalities offered by Web services as generic operations. Community ontology is used as a general template for describing semantic Web services and communities. A major advantage of this work is the peer-to-peer community management solution that addresses the problems of centralized approaches.

In this work, communities gather services from the same domain of interest and publish the functionalities offered by Web services as generic operations. Community ontology is used as a general template for describing semantic Web services and communities. The approach follows a realistic community-centric point-of-view, and adopts a peer-to-peer solution to manage communities, which addresses the problems of centralized approaches.

Metadata ontology, called community ontology is used for creating communities of Web services. Metadata ontologies provide concepts that allow the description of other concepts. Communities are instances of the community ontology. They are created by community providers which use the community ontology as a template. Community providers are generally groups of government agencies, non-profit organizations, and businesses that share a common domain of interest. For example, the Department for the Aging and other related agencies, such as the Department of Health, would define a community that provides healthcare benefits for senior citizens. A community is itself a service that is created, advertised, discovered, and invoked in the same way"regular" Web services are. The providers of a community assign values to the attributes and concepts of the community ontology. Communities are published in registry so that they can be discovered by service providers

Service providers identify the community of interest and register their services with it. A Web service may belong to different communities. For example a composite service may out source operations that have different domains of interest

Since these operations belong to two different communities, the composite service is registered with the "healthcare" and "elderly" communities. End-use selects a community of interest and invokes its operations. Each invocation of a community operation is translated into the invocation of a community member operation.

In [14] authors present a quality-driven approach to select component services during the execution of a composite service. The features of this approach are:

- An extensible multi-dimensional Web service quality model. Dimensions of the model characterize non-functional properties that are inherent to Web services in general: execution price, execution duration, reputation, reliability and availability

- A quality driven service selection: In order to overcome the limitations of local service selection outlined above, authors propose a global planning approach. In this approach quality constraints and preferences are assigned to composite services rather than to individual tasks within a composite service.

A composite service is an umbrella structure aggregating multiple other elementary and composite Web services, which interact with each other according to a process model. The statechart is chosen to specify the process model of a composite service.

A basic state of statechart describing a composite service can be labeled with an invocation to any of the 3 types of services: elementary services, composite services, and service communities defined in SELF-SERV.

The set of members of a community can be fixed when the community is created or it can be determined through a registration mechanism, thereby allowing services providers to join, quit and reinstate the community.

### 3.3. A comparison

We compare the above approaches according to the following requirements as shown in table1:

- Modeling the collaboration: The ability to perform long-lived, peer-to-peer collaboration between participating services. Collaboration must be modeled in terms of interactions of messaging exchanges.

- Semantic support: Web services composition languages should enable the representation of semantics of composed services to facilitate the automated composition of Web services. The semantics descriptions that enable dynamic service discovery and invocation are imperative

- Composition strategy: Four categories of composition strategies have identified here: Declarative, Model-driven, Ontology-based and Context-based Web services composition have been dealt with.

Table 1.  The comparison

| Approach | Semantic support | Collaboration support | Composition strategy |
|---|---|---|---|
| SELF-SERV [10] | No | Yes | Declarative |
| WS-CatalogNET [8] | Yes | Yes | Model-driven |
| WSC[7] | Yes | No | Ontology-based |
| [11] | Yes | Yes | Ontology-based |
| [12] | Yes | Yes | Context-based |
| Community-based architecture[13] | Yes | Yes | Ontology-based |
| quality-driven approach[14] | No | No | Context-based |

## 4. QUERYING WEB SERVICES COMMUNITIES

In fact, satisfying users' queries is in the heart of organizing Web services into communities. In most cases, user's queries require the composition of several Web services belonging to different communities. Different approaches tackle the problem of the query processing on communities

### 4.1 Approaches for Querying Web services communities

In WSCatalogNet [15], queries are processed by the members of a peer, but routing of the queries is a responsibility of the peers. The purpose of the query routing is to identify a set of members that, when put together, can satisfy all constraints required by a query. Hence, a routing takes place before the actual query process. Once a set of members (not necessarily from the same peer) are identified, queries are sent to each member in the set for processing. The results are combined by the original community. Since a community does not store product data locally, processing the query requires locating catalogs that are capable of answering the query. The authors in [16] propose a cooperative query processing technique that consists of two steps:

- Identify best combinations of members whose query capabilities, when put together, satisfy the constraints expressed in the query.

- Resolve the query by sending it to the selected combination of members.

A query rewriting algorithm is developed and adopted by the authors [17]; Best Query Rewriting (BQR) [18]. This algorithm identifies which part of the query can be answered by local members of the community and which part of the query cannot (hence, needs help of peers). The algorithm takes as input the community schema, member definitions and the query (all expressed in the class description language) then produces the following output:

- Qlocal: the part of the query Q that can be answered by the community's local members. It gives the best combinations of the local members that can answer all (or part of) the query.

- Qrest: the part of the query that cannot be answered by the local members. This part of the query will be forwarded to peer communities. It is noted that the expected answers of the forwarding is the combination of the external members (i.e. members of peer communities) that are capable of answering the part of the query.

Each community has a query forwarding policy which controls what should be done with Qrest. The forwarding policy can express (i) when the query should be forwarded (e.g. when no local members can answer, when the community is too busy, etc.) (ii) to which peer (e.g. all, top K, random, etc.) the query should be forwarded, and (iii) how far the query should be forwarded (i.e., hop limit). After forwarding, the community collects the returned results from the peers and chooses the best combination of e-catalog members (local and external) based on the quality of the members' (e.g. reliability) and user preferences. After all necessary members are selected, each of the selected member processes parts of the query that it is capable of processing, and the results are returned to the community.

CONSERV [19] is a middleware infrastructure, which aims at providing context-aware querying of information, provided in a pervasive computing environment that consists of ad-hoc communities of web services. The main objective of the CONSERV architecture [19] is the facilitation of the answering of queries over communities of peers. The cornerstone of the CONSERV architecture is the replacement of the traditional treatment of databases as persistent collections of tuples by the assumption that a database relation is a collection of tuples dynamically compiled from an ad-hoc community of peers, each offering tuples to the relation through a Workflow of web services. Queries are posed against the database of a peer. The "user" that issues a query need only know the names and schemata of the relations being used; the nature of the relations and the workflows necessary for the collection of the values of the virtual relations are transparent to the user. The query is expressed in standard SQL (i.e., the nature of the involved relations is transparent to the user) and the collection of tuples is automatically performed by the system.

The SQL query is parsed by the query processor. As in traditional DBMSs, the query processor receives a declarative SQL query and produces a procedural execution tree to be issued against the underlying data. The execution tree involves the integration of information coming from different peers. There are several steps to be taken towards the construction of the execution tree:

- Identification of the peers to be probed for tuples. To facilitate this task, there is a directory of known peers in the community of the peer serving the question and a peer manager that ultimately determines which peers are to be contacted.

- Identification of the workflows of web services that need to be invoked for each peer. In the simplest case, each relation in the local database is linked to the execution of one or more web services in remote peers. Each of these web services, in turn, returns a message that corresponds to one, several, or all the attributes of the relation that we wish to populate. In more complicated cases, it is quite possible that we need to transform, merge, cleanse or, in any case, process this incoming information before propagating it further towards the local relation.

Hence, a workflow of web service operations is needed in order to obtain the tuples from each peer. The complexity of the workflow may vary along with the overhead introduced during its execution. The determination of this workflow is performed by the workflow resolver. In CONSERV, workflows are treated as connected digraphs comprising at least a fountain start node and a sink end node. The peer's directory, the peer's manager and the workflow resolver form the context manager subsystem, which together with the query processor constitute the overall CONSERV architecture.

In [20] authors propose a novel approach for querying and automatically composing Data providing services. The proposed approach largely draws from the experiences lessons learned in the areas of service composition, ontology, and answering queries over views. First, it introduces a model for the description of Data Providing services and specification of service-oriented queries. Data Providing services are modeled as RDF views over a mediated (domain) ontology. Each RDF view contains concepts and relations from the mediated ontology to capture the semantic relationships between input and output parameters. Second, a query rewriting algorithm is proposed for processing queries over Data Providing services. The query mediator automatically transforms a user's query (during the query rewriting stage) into a composition of DP services. The contributions of this paper are summarized below:

- Query Model for DP Services: 1) An RDF-based model for the description of DP services is proposed and 2) specification of service-oriented queries.DP services are modeled as RDF views over domain ontologies. Input/output relationships are declaratively represented based on concepts and relations that are semantically defined

- in a mediated ontology. SPARQL language is adopted for posing queries over DP services.
- Processing DP Service Queries—a query rewriting algorithms for processing queries over DP services is proposed. The idea behind query rewriting is the following: given a query over the mediated ontology and a set of RDF views of DP services, reformulate the query into an expression that refers only to the RDF views and provides the answer to the query.

The proposed approach automatically transforms a user's query (during query rewriting) into a composition of DP services (modeled as RDF views) that are selected, orchestrated, and invoked to execute the posed query.

### 4.2 Discussion

In fact , the main requirements of the query processing among communities are: the identification of relevant communities that contribute in the generation of the answer, the collection of answers from communities and the delivering of answers to user. In following, we state the different tasks involved in query processing as shown in figure 2:

- Formulating a query: The Users may view the and access information about the community in general, but they are not able to navigate through the system's functionality. Their main role is to request a service.
- The query processing between communities is performed according to the following steps:
- Identifying concerned communities: Identifying the combination of members whose query capabilities, when put together, satisfy all constraints expressed in the query. The members can be local (belonging to the community), or external (belonging to the community peers).
- Rewriting Query for concerned communities: The Query is divided into sub-queries to the identified communities.
- Routing Queries among Communities: The sub-queries are sent to the identified communities.
- identifying communities' members concerned by the query.
- Rewriting the query for concerned members.
- Finally,delivering the results to the user.

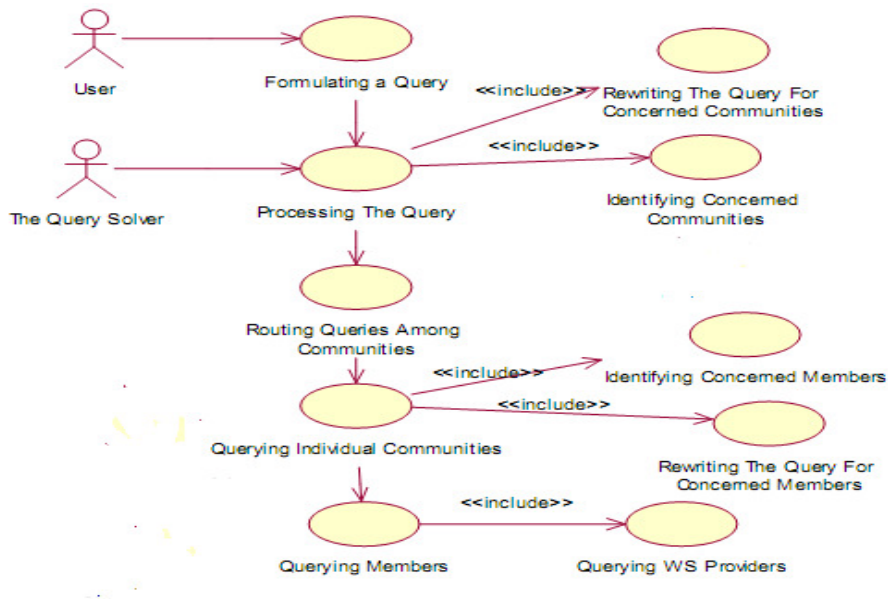

Figure 2. UML Use Case diagram for querying communities

We can say that the described approaches [15], [19] and [20] propose solutions for all the described requirements and the difference is the query language chosen in the query expression. While WSCatalogNet[15] uses the description Logic, CONSERV [19] uses SQL and [20] uses SPARQL.

### 4.3. Web services selection in Web services communities

In fact query processing requires the selection of an appropriate Web service for a particular task which has become a difficult challenge due to the increasing number of Web services offering similar functionalities.

#### 4.3.1. Web services selection approaches

In the following we dress a list of current Web services selection approaches.

In Taher et al.'s work [21], Web service selection is performed according to a set of QoS criteria (speed, reliability, reputation, etc.). The community is also in charge of administrative tasks such as addition and suppression of services to and from the community. Web service substitution is also addressed in this work and consists of replacing a non-functioning or non-responding Web service with a functionally equivalent one, in order to find an alternative way to enable a composition in case of exception.

Web service substitution consists in replacing a disfunctioning or non responding Web service with a functionally equivalent one, in order to find an alternative way to enable a composition in case of exception. Substituting a service within another requires the mediation of communications between the replacing service and the original client. Mediator Web services communicate with the concrete Web services that implement the functionality, each mediator connects to a specific service.

In [22] authors propose a community-based approach for web service selection where super-agents with more capabilities serve as community managers. They maintain communities and build community-based reputation for a service based on the opinions from all community members that have similar interests and judgement criteria. The community-based reputation is useful for consumer agents in selecting satisfactory services when they do not have much personal experience with the services. A practical reward mechanism is also introduced to create incentives for super-agents to contribute their resources and provide truthful community-based reputation information, as strong support for the approach.

A simple Web services selection schema based on user's requirement of the various non-functional properties and interaction with the system is proposed in [23]  . The proposed framework utilizes user preferences as an additional input to the selection engine and the system ranks the available services based on the user preferences. The proposed architecture also relies on selection and matching engines, which interact with service communities.

The Web services selection in virtual communities [24]  follows these steps:

- In the first stage, web services are generally selected from a repository system or marketplace on the basis of its interface description, basically comprising a list of provided methods, and several non-functional  properties such as the geographical location of the  service provider, performance, its price, and so on.

- Web service selection can be performed from two perspectives: bottom-up and top-down. Top-down selection of web services starts from the business processes, e.g., setting up a course, and then identifies those services whose capabilities and quality aspects conform best. The bottom-up perspective, on the other hand, starts from the available web services, and tries to select those that fit best. In practice, both selection approaches are often combined

An approach that shows that  context of ontology may affect the quality of service  selection is proposed in [25].It  made  also a compromise in using  contextual ontology, on a single representation  imposed by the use of QoS ontology and the  multiplicity of local ontology of Web services. The concept of context around which relies the proposed method has several advantages in terms of opportunities that it affords for advances in web service selection. Furthermore, prospects remain open, not only in the field of Web services, but more generally in various fields involving the interoperability of data. Hence, the context of ontology has effects on the degree of match which is the core of the pragmatic selection. Moreover, this approach may be extended to automatic service selection using multi dimensional QoS.

### 4.3.2. Summary

So far, we have dealt with many different approaches that have been developed in order to facilitate Web Services selection among communities.

 The aim of this section is to give a summary of what we have presented so far. This is achieved by creating a table, listing most of the selection approaches    that have been discussed in this paper.

| Approach | Selection Criteria |
|----------|-------------------|
| [21]]    | Set of QoS |
| [22]     | Reputation |
| [23]     | User Preferences |
| [24]     | Interface Description of Web Services |
| [25]     | Context Ontology |

Table 2. Web Services selection approaches

Table 2 provides an overview on current Web Service selection approaches. The reader will also notice that what make the difference between proposed approaches is the criteria on which the selection process is based

We think that the current state of the development efforts can be seen as first attempts to solve the selection problem using additional specifications concerning QoS[21] , [22] user preferences[23] , interface description of Web Services[24] .

Unfortunately, most papers do not provide information about how well the described approach already works. In our view, they lack semantic description, and thus, it would also be interesting to follow the progress offered in [25] with the context-based approach that will be made in the area of semantic web services.

## 5. CONCLUSION

In this paper, we first draw an overview of major Web services communities' management approaches. Current approaches were chosen and compared against requirements that an approach should support to facilitate the management task. Second, the problem of the querying communities was tackled through the main requirements it involves.

In fact, the studied issues allow us introduce the core of our future work :proposing and detailing an architecture which aims to meet Web Services Communities management and querying issues .The purpose of the architecture is to take advantages of the studied approaches and in a optimized way .

## REFERENCES

[1]   Lee, S.hyun. & Kim Mi Na, (2008) "This is my paper", ABC *Transactions on ECE*, Vol. 10, No. 5, pp120-1

[2]   Gizem, Aksahya & Ayese, Ozcan (2009) *Coomunications & Networks*, Network Books, ABC Publishers.

[3]   Kozinets, R. V, (1999) "The Strategic Implications of Virtual Communities of Consumption", *European Management Journal*, Vol. 17, No.3, pp.252-264.

[4]   Wellman, B, (2001) "Computer Networks as Social Networks", *Science 293*, pp2031-2034.

[5]   Smith, M, (2002) "Tools for Navigating Large Social Cyberspaces", *Comm. of the ACM 45(4)*, pp 51-55.

[6]   Schubert, P. & M. Ginsburg, (2000) " Virtual Communities of Transaction: The Role of Personalization in Electronic Commerce", *Electronic Markets 10(1),* pp 45-55.

[7]    I. Foster, T. Freeman, K. Keahey, D. Scheftner, B. Sotomayer, &  X. Zhang (2006) "Virtual Clusters for Grid Communities" , *Sixth IEEE International Symposium on Cluster Computing and the Grid, Singapore, IEEE Computer Society*, pp 513–520.

[8]    H. Paik, B. Benatallah, & R. Hamadi, (2002) "Dynamic Restructuring of E-Catalog Communities Based on User Interaction Patterns" WWW *Journal*, pp 325–366.

[9]    H.Xuejuan , C.Xinmeng , &  L. Jinshuo, (2008) "Dynamic Selection of Best Service in Semantic Web Service Community", *Proceedings of the International Conference on Computer Science and Software Engineering,* Vol. 03 , pp 407-410.

[10]   H. Paik, B. Benatallah. & F. Toumani, (2004) "WS-CatalogNet: Building Peer-to-Peer e-Catalog*", In Proc. of 6th International Conference on Flexible Query Answering Systems.*

[11]   Benatallah,B Dumas,M , &   Sheng, Q Z. (2005) "Facilitating the Rapid Development and Scalable Orchestration of Composite Web Services". *Distributed and Parallel Databases*, Vol 17. pp. 5-37.

[12]   Sheng, Q.Z., Benatallah, B., Dumas, &   M. Mak, E.O-Y. (2002) "SELF-SERV: a platform for rapid composition of web services  in a peer-to-peer environment", *Proceedings of the  28th VLDB Conferen*ce, Hong Kong, China.

[13]   S. Sattanathan,  P. Thiran, Z. Maamar . & D. Benslimane, (2007) " Engineering communities of web services", iiWAS  *Austrian Computer Society*, Vol.229, pp57-66.

[14]   M. Mrissa, C. Ghedira, D. Benslimane, Z. Maamar, F. Rosenberg, S. Dustdar, (2007) "A context-based mediation approach to compose  semantic web services". *ACM Trans. Internet Techn*.

[15]   B. Medjahed . &   A. Bouguettaya (2005) " A dynamic foundational architecture for semantic web services",  *Distributed and Parallel Databases*,  Vol. 17, No. 2, pp179-206.

[16]   L. Zeng , B. Benatallah, M, Dumas, J. Kalagnanam, &  Q.Z. Sheng,(2003) "Quality driven Web services composition*", in: Proceeding of the 12th International World Wide Web Conference (WWW2003)*, Budapest , Hungary .

[17]   H.-Y. Paik, B. Benatallah &  F. Toumani  ,(2005) "Towards Self-Organizing Service Communities".*IEEE  Transactions on Sys. Man and Cyb* , Vol.35 , pp. 408-419 .

[18]   B. Benatallah, M.-S. Hacid, H. Paik, C. Rey &F.Toumani(2004) "Towards Semantic-driven, Flexible and Scalable Framework for Peering and Querying e-Catalog Communities". *Information Systems Journal,Special issue on semantic web services*.

[19]   K. Baina, B. Benatallah, H. Paik, F. Toumani, C. Rey, A. Rutkowska. & B. Harianto, (2004) "WS-CatalogNet: An Infrastructure for Creating, Peering, and Querying e-Catalog Communities" ,  *In Proc. of the 30th VLDB Conference.*

[20]   M. J. Carey, M. J. Franklin, M. Livny E. &  J. Shekita, (1991) "Data Caching Tradeoffs in Client-Server" DBMS Architectures,  In Proc. SIGMOD *Conference*, pp  357-366.

[21]   A. Zarras, P. Vassiliadis, &  E. Pitoura,(2005) "Query management over ad-hoc communities of web services", in: Proceedings of the 2nd IEEE International Conference on Pervasive Services, pp. 261-270

[22]   D. Benslimane, M. Barhamgi. & M. Medjahed, (2010) "A Query Rewriting Approach for Web Service Composition",  IEEE *transactions on service computing*, Vol.3.

[23]     Y. Taher, D. Benslimane, M.-C. Fauvet & Z. Maamar, (2006) " Towards an approach for web services Substitution", *IOS Press*, pp 166–173.

[24]     Y.Wang, J. Zhang. & J.Vassileva, (2010) "Effective Web Service Selection via Communities Formed by Super-Agents", IEEE / WIC / ACM *International Conferences on Web intelligence.*

[25]     Y. Badr, A. Abraham, F.Biennier & C. Grosan (2008) "Enhancing Web Service Selection by User Preferences of Non-Functional Features", *Fourth International Conference on Next Generation Web Services Practices* .

[26]     A. Moor, V. Heuvel, (2004) "Web service selection in virtual communities," *System Sciences Proceedings of the 37th Annual International Conference*

[27]     N. Keskes, A. Lehireche & A.Rahmoun (2010) "Web Services Selection Based on Context Ontology and Quality of Services" *International Arab Journal of e-Technology*, Vol. 1, No. 3